\documentclass[10pt]{article}

\usepackage{epsfig}

\newlength{\dinwidth}
\newlength{\dinmargin}
\marginparpush
5pt \topmargin -42pt \headheight 12pt

\newcommand{\aver}[1]{\langle \, #1 \, \rangle \mathstrut}
\newcommand{\fr}[2]{\!\left(\!\frac{#1}{#2}\!\right)}
\newcommand{\fra}[3]{\!\left(\!\frac{#1}{#2}\!\right)^{\!#3}}

\newcommand{\keV}{\, \mathrm{keV}}

\newcommand{\GeV}{\, \mathrm{GeV}}

\def\({\left(}
\def\){\right)}
\def\cm{{\,\rm cm}}

\begin{document}
\titlepage

\vspace*{4cm}

\begin{center}

{\Large \bf $^3$He {\it experimentum crucis} for Dark Matter puzzles}

\vspace*{1cm}
\textsc{K. Belotsky$^{a,b}$, Yu. Bunkov$^{c}$,
H. Godfrin$^{c}$, M. Khlopov$^{a,b,c}$, R. Konoplich$^{d}$} \\

\vspace*{0.5cm}
$^a$ Moscow Engineering and Physics Institute (State University), \\ Kashirskoe
shosse 31, 115409,
Moscow, Russia \\[0.5ex]
$^b$ Centre for Cosmoparticle Physics "Cosmion", \\
4 Miusskaya pl., 125047, Moscow, Russia \\[0.5ex]
$^c$ Centre de Recherches sur les Tr\`es Basses Temp\'eratures, CNRS, \\
Laboratoire associ\'e \`a l'Universit\'e J.Fourier et ˆ l'Institut National Polytechnique de Grenoble \\
25, avenue des Martyrs,  38042, Grenoble Cedex 09, France \\[0.5ex]
$^d$ Department of Physics, New York University,\\
New York, NY 10003, USA\\
Department of Physics, Manhattan College, Riverdale, New
York, NY 10471, USA

\end{center}

\vspace*{1cm}

\begin{abstract}

The leading direct dark matter search experiments: CDMS, Edelweis
and DAMA/NaI exhibit different results for different approaches to
the problem. This contradiction can reflect a nontrivial and probably a
multi-component nature of the cosmological dark matter. WIMPs can
possess dominantly a Spin Dependent interaction with nucleons. They
can be superheavy or represent atom-like systems of superheavy
charged particles. The Dark matter can contain a component, which
strongly interacts with the matter. We show that even a moderate size
superfluid $^3$He detector provides a crucial test for these
hypotheses and that its existing laboratory prototype is already of interest for 
the experimental dark matter search.
\end{abstract}

\section{Introduction}

The existence of nonbaryonic dark matter (DM) is one of the cornerstones
of modern cosmology. Possible physical candidates for DM particles correspond to different extensions
of the Standard Model, describing the known sector of particle physics. Theoretical reasons for such extensions
are sufficiently serious to consider these candidates altogether and to expect a multi-component nature of
Dark Matter.
Recently developed methods of precision cosmology provide the possibility to reveal the dominant form of
Dark Matter in astrophysical observations. However, these methods alone are insufficient to explore
the problem in
all its complexity, so the set of different experimental probes should be developed to distinguish
various DM components
(including subdominant ones) by their specific effects.

An important step in the exploration of the DM problem was the
development of direct searches for Weakly Interacting Massive
Particles (WIMP). Underground detectors were created in which
effects of nucleus recoil induced by interaction of cosmic WIMP with
a nucleus were searched for. As a result of this search 
DAMA/NaI underground experiment claimed
the observation of annual modulation signature \cite{DAMA-review}.
The Si and Ge low temperature experiments with a high level of
discrimination did not find the dark matter signals at the 
corresponding level of probability \cite{CDMS,Edelw}. The reasons of
this discrepancy are now under hot discussion. In particular, this
discrepancy can reflect the existence of non-dominant DM component in
the form of heavy neutrinos of the 4th generation \cite{Fargion,BK}.
A possibility to explain the DAMA/NaI events within the framework of
the Standard Model extended to the 4th generation of fermions,
described in \cite{Fargion}, was recently confronted with the
results of CDMS and Edelweiss and it was found that this hypothesis
can provide an explanation for all these results. It was also shown
in \cite{machet-plb} that a wide class of supersymmetric models can
also possess this property. It may be a class of DM models of purely
Spin Dependent (SD) interaction of WIMPs with nuclei. Another
solution can be related with superheavy dark matter particles, which
are slowed down in terrestrial matter below a detection threshold of
CDMS and Edelweiss. Slowing down assumes strong interaction with
matter and the interaction can be so strong that the slowed down
particles can not cause any effect above threshold of underground
detectors. For such Strongly Interacting Massive Particles (SIMPs)
\cite{Starkman} a very small size X-ray detector during only 100
seconds of XQC experiment \cite{XQC,XQC2} provided severe
constraints. However, the high sensitivity of XQC experiment in
tens-eV range is completely lost, when energy release approaches keV
range, and there is a gap in experimental sensitivity to energy
release in the interval from 1 to 6-10 keV. Moreover, it turns out
that the results of XQC experiment are practically insensitive to
the existence of such a dark matter candidate, as the recently proposed
nuclear-interacting O-helium \cite{anutium}.

Superfluid $^3$He-B at ultra-low temperatures is an appealing target
material for bolometric particle detection \cite{First,Second,BOLO},
and particularly for the search for non-baryonic Dark Matter. The
main arguments in favor of $^3$He are its non-zero nuclear magnetic
moment (allowing therefore to explore the Spin-Dependent interaction
channel) combined to the extremely high sensitivity of superfluid
$^3$He bolometers and the possibility of efficient neutron
background discrimination. It is the purpose of the present paper to
specify the class of DM models, for which even moderate size
superfluid $^3$He bolometers can provide a crucial experimental
test.

To minimize the number of free parameters for SD interacting DM we
follow the phenomenological framework \cite{DAMA-review}, which
provides direct correlation between the results of the DAMA/NaI
experiment and the expected signal in superfluid $^3$He detectors.
Our approach is aimed to study sensitivity of such detectors in a
model independent way and is complementary to similar studies for
particular models, e.g. for supersymmetric models
\cite{machet-plb,bednyakov,Bednyakov:2003wf}. We also consider a
wider class of dark matter models, involving, in particular,
superheavy particles, strongly interacting with matter. The account
for nontrivial forms of dark matter, which are not reduced to WIMPs,
may be crucial in the resolution of current dark matter puzzles.

\section{$^3$He response to SD interacting Dark Matter}

The results of the analysis of DAMA data for a combination of
Spin-Independent (SI) and Spin-Dependent (SD) interaction of WIMPs
with matter are given in Fig. 30 of reference \cite{DAMA-review}.
These results are usually presented in terms of the product $\xi
\sigma$ of the relative contribution $\xi$ of WIMP density
$\rho_{X}$ into the total local halo density $0.17 <\rho_{loc} <
1.7$GeV/cm$^3$
\begin{equation}
\xi = \frac{\rho_{X}}{\rho_{loc}}
\label{xi}
\end{equation}
and the cross sections $\sigma$ of SI and SD interactions.

The values of $\xi \sigma_{SI}$ and $\xi \sigma_{SD}$ differ for SI and SD interactions by 5 orders of magnitude,
what provides plausible reasons for testing the purely SD-interacting WIMP nature of DAMA events even in a relatively
modest size superfluid $^3$He detector.

\subsection{Qualitative estimation}

We first give rough estimation of the expected number of events in a detector containing $N_m$ moles
of superfluid $^3$He.

Let $n= \xi \frac{\rho_{loc}}{m_X}$ be the local number density of WIMPs with mass $m_X$ and
galactic averaged velocity $v$ ($v \sim 300$ km/s).
For the cross section $\sigma_{SD}$ the number of events $N_{ev}$ expected in the detector during the period
$T \sim 1$ year of its operation is given by
\begin{equation}
N_{ev}=n \sigma_{SD} v T N_A N_m,
\label{ev}
\end{equation}
where $N_A=6 \cdot 10^{23}\,{\rm mole^{-1}}$ is the Avogadro number.

We can estimate the minimal number of events, corresponding to the lower limits on
$\xi \sigma_{SD}$ in \cite{DAMA-review}.
This minimal estimation corresponds to maximal values of the astrophysical parameter
$\rho_{loc} =1.7$ GeV/cm$^3$, so that $n=3.4 \cdot 10^{-2} \cdot \xi \cdot (\frac{50 \GeV}{m_X})$ cm$^{-3}$.

For this minimal estimation one obtains from Eq.(\ref{ev})
$$N_{ev} = 18 \frac{(\xi \sigma_{SD})_{min}}{1\,\rm pb} N_m \fr{v}{300\, \rm km/s} \fr{T}{1\, \rm yr} \fr{50\GeV}{m_X}.$$

Since in the whole range of SD-interacting WIMP parameters, reproducing the DAMA result, the value of
$(\xi \sigma_{SD})_{min} > 10^{-2} $ pb, $v > 100$ km/s, $m_X < 110$ GeV, one finds that a detector containing
$N_m \ge 30$ moles of superfluid $^3$He can cover all the possible range of these parameters during
one year of operation.
It would provide a complete experimental test for the SD-interacting WIMP interpretation of the DAMA event.
However, this general optimistic estimation should be taken with caution in view of the theoretical uncertainty
in the possible parameters and nature of the SD interaction.

\subsection{Phenomenology of SD interacting WIMPs}

In the non-relativistic limit, which is appropriate for WIMPs in the Galaxy,
the variety of possible forms of
WIMP-nucleus interaction
is reduced to two cases, namely, to a spin-spin interaction and to a scalar one.
Fundamental constants of WIMP interaction with nucleon constituents, being specified by each concrete particle model,
determine the effective coupling of WIMPs to nucleons, which, in turn, defines constants of WIMP-nucleus interaction
(for a more detailed review see \cite{Kamionk,Goodman,KurKam}).
The essential difference between spin-spin and scalar interactions is in the following.
In the scalar case, the WIMP-nucleus  interaction amplitude ($A_{XA}$) is given by the WIMP-nucleon ($A_{Xp,n}$) amplitude,
multiplied by the number of respective nucleons, while in the spin-spin case $A_{XA}$ is proportional
to the nucleon spin averaged over the nucleus state $\aver{S_{p,n}}$, which for heavy non-zero spin nuclei is,
as a rule, even smaller than that for a single nucleon ($S_p=S_n=1/2$).
It leads to a loss of advantage in using heavy target-nuclei in the exploration of WIMPs with spin dependent interaction.

Let us denote, according to
\cite{DAMA-review,Kamionk}, $a_p$ and $a_n$ the coupling constants of WIMP SD-interaction with proton and neutron,
respectively. Then the cross section of SD interaction between WIMP and a nucleus with spin $J$
can be represented as \cite{DAMA-review}
\begin{equation}
\sigma_{SD}=\fra{\mu_{XA}}{\mu_{Xp}}{2}\frac{4}{3}\frac{J+1}{J}\sigma_{SD}^{(np)}
\left(\aver{S_p}\cos{\theta}+\aver{S_n}\sin{\theta}\right)^2G_{SD}.
\label{sigmaSD}
\end{equation}
Here, following \cite{DAMA-review},
a parameter $\theta=arctg\frac{a_n}{a_p}$ is introduced to characterize the relative contribution to WIMP-nucleus interaction
from WIMP-proton and WIMP-neutron couplings. In (\ref{sigmaSD}) $\mu_{Xp}$ and $\mu_{XA}$ are the reduced masses
of WIMP ($m_X$) and proton (nucleon) ($m_p$) and nucleus ($m_A$),
\begin{equation}
\sigma_{SD}^{(np)}=\frac{32}{\pi}\frac{3}{4}G_F^2\mu_{Xp}(a_p^2+a_n^2)
\end{equation}
being denoted in \cite{DAMA-review} as $\sigma_{SD}$,
\begin{equation}
G_{SD}=\frac{1}{E_{R\max}}\int_0^{E_{R\max}} F_{SD}^2(E_R)dE_R
\end{equation}
takes into account effects of the finite size of the nucleus (a loss of coherence). 
The effect of the finite size of nucleus is conveniently parametrized by dimensionless $y=\fr{|\vec{q}|b}{2}^2=\frac{m_AE_Rb^2}{2}$ 
with $\vec{q}$ being transferred momentum and $b=1\,{\rm fm}\,A^{1/6}$
(or more precisely \cite{Ressell} $b=\sqrt{\frac{41.467}{45A^{-1/3}-25A^{-2/3}}}$ fm). For $y \ll 1$, $G_{SD} =1$. 
The representation of $\sigma_{SD}$ by Eq.(\ref{sigmaSD}) implies normalization of the form-factor $F_{SD}$ on unity at
zero transferred energy $E_R$
\begin{equation}
F^2_{SD}=\frac{S(E_R)}{S(0)}.
\label{FSD}
\end{equation}
The function $S(E_R)$ is conventionally divided in
isovector ($S_{11}$), isoscalar ($S_{00}$) and interference ($S_{01}$) parts and has the form
\begin{equation}
S(y)=(a_p+a_n)^2S_{00}(y)+(a_p-a_n)^2S_{11}(y)+(a_p^2-a_n^2)S_{01}(y).
\end{equation}
Since $a_p$ and $a_n$ directly enter this expression,
$S$ depends on the type  of WIMP.
The larger is the dimensionless argument $y$, the stronger is the suppression
due to finite nucleus size. $F_{SD}$, being defined in the form of Eq.(\ref{FSD}),
depends on the WIMP type ($a_{p,n}$) only through $\theta$ (which in first approximation can be neglected
for $^{23}$Na and $^{127}$I, used in DAMA/NaI setup \cite{Tovey}). It makes the value $\sigma_{SD}^{(np)}$ inferred
from DAMA measurements (see Eq.(\ref{sigmaSD})) less WIMP model-dependent at fixed $\theta$.

Eventually one notes, that Eq.(\ref{sigmaSD}) represents SD WIMP-nucleus cross-section through three
WIMP-type dependent parameters: $\sigma_{SD}^{(np)}(a_p,a_n)$, $\theta(a_p/a_n)$ and $m_X$.
The values $\aver{S_{p,n}}$ and functions $S_{ij}(y)$ are defined by
a particular nuclear model and suffer with considerable
uncertanties \cite{DAMA-review,Kamionk,Ressell}.

\subsection{The expected event rates in $^3$He}

There is an important difference between the nucleus $^3$He and nuclei $^{23}$Na and $^{127}$I, used in DAMA/NaI set-up.
$^3$He has an unpaired neutron and is an even-odd nucleus (in terms of numbers of its protons and neutrons).
As a consequence, the spin of this nucleus is determined mainly by
the neutron\footnote{In principle, this statement does not mean that $\aver{S_n}=J$,
because nucleons' orbital momentum in most cases essentially contributes into $J$ too.
However, this is not the case for $^3$He.} $\aver{S_p}\approx 0$, $\aver{S_n}\neq 0,$
while $^{23}$Na and $^{127}$I are odd-even and for them $\aver{S_p}\neq 0$, $\aver{S_n}\approx 0$.
Deviation of $\aver{S_p}$ from zero for even-proton nuclei (and analogously for even-neutron) is determined
by the details of nuclear model \cite{Ressell}. But for the "simply" composed $^3$He nucleus ($J=1/2$)
it will be quite accurate to use the single particle (or also odd group) shell model of nucleus
which gives here $\aver{S_p}=0$, $\aver{S_n}=1/2$ \cite{Kamionk}.

The effect of finite size of $^3$He is insignificant. Estimation of the maximal magnitude of $y$ shows it.
The maximal transferred (recoil) momentum for an incident WIMP with velocity $v$ is $|\vec{q}_{\max}|=2\mu_{XA}v$.
Taking into account that $\mu_{XA}<\min\{m_A, m_X\}\le m_A$,
$v<700$ km/s one obtains for $^3$He $y<1.6\cdot 10^{-3}\ll 1$. So we will reasonably assume $G_{SD}=1$.

WIMP-$^3$He SD cross section can be written as
\begin{equation}
\sigma_{SD}=\fra{\mu_{XA}}{\mu_{Xp}}{2}\sigma_{SD}^{(np)}\sin^2{\theta}.
\end{equation}
For a given $\sigma_{SD}$ the event rate in a $^3$He setup will be
\begin{equation}
Rate=\frac{\xi \rho_{loc}}{m_X} \sigma_{SD} \bar{v} N_A N_m,
\end{equation}
where $\bar{v}$ is the mean WIMP velocity, which exceeds the
threshold value, corresponding to the minimal energy release
($E_{Rmin}=1\keV$) in $^3$He setup. Possible dependence of $\xi
\sigma_{SD}^{(np)}$ on $m_X$ has been deduced from the analysis of
positive results of DAMA measurements for different $\theta$,
taking into account uncertainties in $\rho_{loc}$ and WIMP
velocity distribution (i.e. $\bar{v}$). As a first approxiamtion
in the estimation of the expected rate on the basis of DAMA data,
we will fix $\rho_{loc}=0.3$ GeV/cm$^3$, $\bar{v}=250$ km/s for
all plots of DAMA (all boundaries enclosing an allowable region).
This simplification prevents double account of uncertanties in
$\rho_{loc}$ and $\bar{v}$ (first one in $\xi \sigma_{SD}^{(np)}$
and second one in $Rate$).

Expected rates per 1 g of superfluid $^3$He, obtained on the basis of DAMA data, are shown on Fig.1 for two cases:
$\theta=\pi/2$ (Fig.30c of \cite{DAMA-review}) and $\theta=2.435$ (Fig.30d of \cite{DAMA-review}).
The case $\theta=\pi/2$ corresponds to WIMP coupling to neutrons only
($a_n\neq 0$, $a_p=0$), what is the most preferable case for the $^3$He setup.
The case $\theta=2.435$ ($a_n/a_p=-0.85$) takes place for the neutralino with purely $Z^0$-boson mediated interaction with nuclei.
The case of $\theta=0$ (WIMPs interact with only protons) is virtually insensitive for a $^3$He setup.
Note that in a $^3$He setup containing 100 g of superfluid $^3$He the rate of event should exceed one events per month.
\begin{figure}[ht]
\begin{center}
\centerline{\epsfxsize=12cm\epsfbox{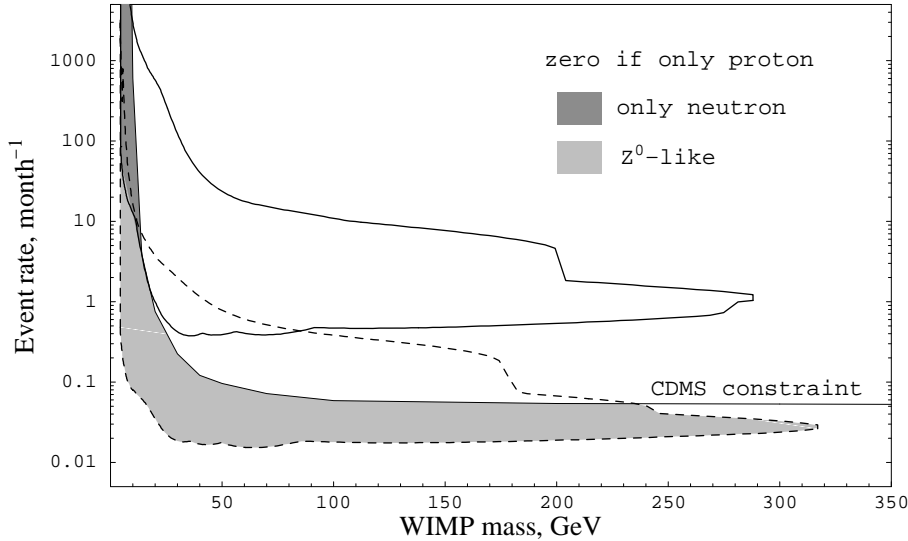}}
\caption{Expected signal in 1 g of superfluid He-3 detector for the interpretation of DAMA/NaI results
in terms of purely Spin Dependent interaction.
Two regions, enclosed by solid and dash lines, correspond to expected event rates
for WIMP-nucleon coupling parameters $\theta=\pi/2$ and $\theta=2.435$ respectively
(see text). CDMS constraint on WIMP-neutron SD cross section is taken into account (thin solid line);
it reduces the expected rate region based on DAMA/NaI as shown by shading.}
\label{DAMAHe1}
\end{center}
\end{figure}

Experiments CDMS and Edelweiss are virtually insensitive to WIMP-proton SD interaction ($a_p$) too. Natural Ge, used in them,
contains mainly spinless isotopes and one odd-neutron isotope $^{73}$Ge. Therefore CDMS and Edeleweiss
are sensitive to WIMP-neutron SD interaction ($a_n$) but suppressed in proportion to the small abundance
fraction of $^{73}$Ge in $^{nat}$Ge ($\sim 8\%$) \cite{CDMS} (In the detector of CDMS II experiment, being under run, Si 
is also used  \cite{Akerib:2005}) . However, the sensitivity of DAMA/NaI to $a_n$ is suppresed too,
because for (even-neutron) $^{127}$I $(\aver{S_n}/\aver{S_p})^2\sim 1/30$ (for $^{23}$Na it is even smaller).
It makes results of CDMS and
Edelweiss important for the exploration of the allowed parameter range for SD interacting WIMPs \cite{KurKam,SIMPLE}.
Some information about this range can be provided by other experiments. Zeplin \cite{Zeplin} and DAMA/Xe,
based on Xe containing odd-neutron isotopes, are sensitive to $a_n$,
NAIAD(NaI) \cite{NAIAD} and CRESST-I(Al) \cite{CRESST} are sensitive to $a_p$, 
while SIMPLE(C$_2$ClF$_5$) \cite{SIMPLE} is sensitive to $a_p$ and less sensitive to $a_n$.
In particular, CDMS data might impose strong constraints on WIMP-neutron SD cross section $\sigma_{Xn}(a_p=0)$
\cite{Gondolo,CDMS}.
For illustration, we took into account the constraint \cite{CDMS} in estimation of the expected event rate in $^3$He set up
(see Fig.\ref{DAMAHe1}). The value $\sigma_{Xn}$, limitted by CDMS at $a_p=0$ ($\theta=\pi/2$),
in terms of introduced notation is $\sigma_{Xn}=\sigma_{SD}^{(np)}\sin^2{\theta}$, entering directly the event rate for $^3$He set up.
In our estimation we treated the limit on $\sigma_{Xn}(\theta=\pi/2)$ as the limit
on this magnitude at any $\theta$. Such treatment neglects the contribution of the WIMP-proton interaction
which is important for target nuclei of CDMS at small $\theta$ and neglects soft $\theta$-dependence of $G_{SD}$ \cite{Tovey}.

Unfortunately, due to the lack of complete experimental details, it is not possible for us to reanalyze all existing experiments
to infer what can be expected using $^3$He.
We therefore concentrate on the DAMA/NaI results, in particular on the nature of the positive signals recorded
by their experiment.

However it is important to note that the low WIMP mass range, where the 
positive signal of DAMA/NaI is near its energy threshold, is
almost inaccessible for probing by other experiments, whereas it
can be fully covered by the He-3 experiment even for an extremely small contribution due to $a_n$.

\section{Multi-component Dark Matter and He-3 detector}
There are a few severely constrained possibilities \cite{KurKam} to explain the 
discrepancy in the 
results of  underground experiments.

For instance, DM particles can be so slow that they are inaccessible for detection by CDMS while
accessible for DAMA. These particles can have a non-standard distribution in the neighborhood of the Solar system, being slow
with respect to it initially. In the latter case, annual modulations remain, but the effect 
goes below the energy threshold of other detectors.

One can suppose that a DM particle possesses such a strong
interaction that it loses its energy going to the detector through
matter (atmosphere and rock) to be able to produce an effect
inside the detector of DAMA/NaI just in a given range, and not
able to do that in CDMS and Edelweiss, being below their
thresholds. For an estimation we assume that $m_X\gg m_A$ for all
nuclei constituting the detectors (note, that $m_X$ relates here
to SIMP - Strongly Interacting Massive Particle). In this case,
the recoil energy of the nucleus lies in the range $0\ldots
2m_Av^2=4\frac{m_A}{m_X}E$, where $v$ and $E$ are the SIMP initial
velocity and energy respectively, and the energy loss of SIMP in
each similar collision is $2m_A/m_X E$.

One can pick out two conditions for such SIMPs to be viable
candidates accounting for the results of underground experiments.
First, SIMPs must lose sufficient energy before reaching the
detectors. Second, the recoil energy spectrum in DAMA must be
ranged below 10 keV. It gives the range of possible masses and
cross sections for SIMPs, which will be accessible to a superfluid
$^3$He detector.

In the multi-component dark matter framework the coexistence of
different components implies experimental means of their
discrimination. In such general context various types of DM
detectors can be complementary and sensitive to different DM
components. In particular, taking into account possible non-WIMP
interpretation of DAMA/NaI data in terms of light scalar bosons
\cite{Bernabei:2005ca} or various multi-component scenarios, such
as \cite{Mitra:2004hh}, one should accept a more general approach
and investigate the efficiency of a He3 set-up to various DM candidates
without normalization on underground experiment data. We
illustrate this approach by a successive discussion of possible
test for composite dark matter models.

\subsection{Composite dark matter and its species}

The recently proposed "sinister" $SU(3)_c \times SU(2) \times SU(2)'
\times U(1)$ gauge model \cite{Glashow} offers an interesting
realization for superheavy WIMPs. It involves three heavy
generations of tera-fermions, which are related with ordinary
light fermions (quarks and leptons) by $CP'$ transformation
linking light fermions to charge conjugates of their heavy
partners and vice versa. $CP'$ symmetry breaking makes
tera-fermions much heavier than their light partners. Tera-fermion
mass pattern is the same as for light generations, but all the
masses are multiplied by the same factor $S > 2 \cdot 10^5$.
Strict conservation of $F = (B-L) - (B'-
 L')$ prevents mixing of charged tera-fermions with light quarks
and leptons. Tera-fermions are sterile relative to SU(2)
electroweak interaction: they do not interact with W, Z and Higgs bosons and thus do not contribute to standard model parameters.
That is why precise measurement of these parameters puts no constraints on properties of tera-particles.
In such realization, the new heavy neutrinos ($N_i$)  acquire large masses and their
mixing with light neutrinos $\nu$ provides a "see-saw" mechanism of
light neutrino Dirac mass generation. Therefore in a Sinister model the
heavy neutrino is unstable. On the contrary, in this scheme E is
the lightest heavy fermion and it is absolutely stable.

In the "Sinister" scenario
very heavy quarks $Q$ (or antiquarks $\bar Q$) can form bound states with other heavy quarks (or antiquarks)
due to their Coulomb-like QCD attraction, and the binding energy of these states may substantially exceed
the binding energy of QCD confinement.
Then $(QQq)$ and $(QQQ)$ baryons must exist.
In the model \cite{Glashow} the properties of heavy generation
fermions are fixed by their discrete $CP'$ symmetry with light
fermions. According to this model a heavy quark $U$ with mass $m_U=S \cdot m_u=S \cdot 3.5 MeV$ and heavy
electron $E$ ($m_E=S \cdot m_e=S \cdot 0.5 MeV$) are stable and
can form a neutral and strongly bound $(UUUEE)$ "atom"
with $(UUU)$ hadron (spin 3/2) as nucleus and two $E$s as "electrons". The
tera gas of such "atoms" can be a candidate for dark matter.

There is an uncertainty in the estimation of cross section of tera-helium interaction with nuclei.
Hadronic interaction of $(UUU)$ with ordinary  nucleons is strongly suppressed, due to a very small size
of this "baryon", which can not be resolved by gluons from ordinary baryons.
The minimal estimation comes from the interaction of the $(UUU)$ magnetic moment with the charge of nucleus.
The cross section of spin 3/2 particle with magnetic moment on a
point-like fermion of charge $eZ$ in the non-relativistic limit
is given by
\begin{equation}
\sigma = \frac{5}{36\pi} \mu^2 e^2 Z^2 (1 + \log{\left(\frac{2MV^2}{E_{1min}-M}\right)}).
\end{equation}
Here $\mu$ is the magnetic moment, $M$ the mass of nucleus, $V$ the speed
of dark matter particles in units of $c$, $E_{1min}$ the detector
threshold for recoil nuclei.
The cross section can be estimated as
\begin{equation}
\sigma = \frac{5}{36\pi} \mu^2 e^2 Z^2 \sim Z^2 (g/2)^2 2.4\cdot 10^{-34} \left(\frac{m_p}{m_X}\right)^2 cm^2,
\end{equation}
where $g$ is the Lande factor ($g=4$ for spin 3/2), $m_X$ the
mass of terahelium particle ($m_X > 2.3 TeV$), $m_p$ the proton
mass.

The problem of such scenario is inevitable presence of "products
of incomplete combustion" and the necessity to decrease their
abundance. Indeed in analogy to D, $^3$He and Li relics that are
the intermediate catalyzers of $^4$He formation in Standard Big
Bang Nucleosynthesis (SBBN) and are important cosmological tracers
of this process, the tera-lepton and tera-hadron relics from
intermediate stages of a multi-step process of towards a final
$(UUUEE)$ formation must survive with high abundance of {\it
visible} relics in the present Universe. To avoid this trouble an
original idea of $(Ep)$ catalysis was proposed in \cite{Glashow}:
as soon as the temperature falls down below $T\sim I_{Ep}/25 \sim
1 \keV$ neutral $(Ep)$ atom with "ionization potential"
$I_{Ep}=\alpha^2m_p/2=25 \keV$ can be formed. The hope was
\cite{Glashow} that this "atom" must catalyze additional effective
binding of various tera-particle species and to reduce their
abundance below the experimental upper limits.

Unfortunately, as it was shown in \cite{Fargion:2005xz}, this
fascinating picture of Sinister Universe can not be realized.
Tracing in more details cosmological evolution of tera-matter and
strictly following the conjecture of \cite{Glashow}, the troubles
of this approach were revealed and gracious exit from them for any
model assuming -1 charge component of composite atom-like dark
matter was found impossible.

The grave problem is that ordinary $^4$He formed in Standard Big
Bang Nucleosynthesis binds at $T \sim 15 keV$ virtually all the
free $E^-$ into positively charged $(^4HeE^-)^+$ "ion", which puts a 
Coulomb barrier for any successive $E^-E^+$ annihilation or any
effective $EU$ binding. It happens {\it before} the $(Ep)$ atom can be
formed and $(Ep)$ atoms can not be formed, since all the free $E$
are already imprisoned by a $^4$He cage. It removes the hope
\cite{Glashow} on $(Ep)$ atomic catalysis as {\it panacea} from
unwanted tera-particle species. The huge frozen abundance of
tera-leptons in hybrid tera-positronium $(eE^+)$
 and hybrid hydrogen-like tera-helium atom $(^4He Ee)$ and
 in other complex anomalous isotopes can not be removed \cite{Fargion:2005xz}.

In spite of this grave problem the idea of Glashow's Sinister
Universe was very inspiring, and composite dark matter scenarios,
avoiding this trouble, were developed.

The AC-model \cite{5} appeared as realistic elementary particle
model, based on the specific approach of \cite{book} to unify
general relativity, quantum mechanics and gauge symmetry.

This realization naturally embeds the Standard model, both
reproducing its gauge symmetry and Higgs mechanism, but to be
realistic, it should go beyond the standard model and offer
candidates for dark matter. Postulates of noncommutative geometry
put severe constraints on the gauge symmetry group, excluding in
this approach, which can be considered as alternative to
superstring phenomenology, supersymmetric and GUT extensions. The
AC-model \cite{5} extends the fermion content of the Standard
model by two heavy particles with opposite electromagnetic and
Z-boson charges. Having no other gauge charges of the Standard model,
these particles (AC-fermions) behave as heavy stable leptons with
charges $-2e$ and $+2e$, called $A$ and $C$, respectively.
AC-fermions are sterile relative to $SU(2)$ electro-weak
interaction, and do not contribute to the standard model
parameters. In the absence of AC-fermion mixing with light
fermions, AC-fermions can be absolutely stable.
The lower limit for the mass of AC-fermions
tfollows from absence of new charged leptons in LEP. It
was assumed in \cite{FKS,Khlopov:2006uv} that $m_A = m_C = m = 100
S_2{\GeV} $ with free parameter $S_2 \ge 1$.

 Primordial excessive negatively charged $A^{--}$ and positively
charged $C^{++}$ form a neutral most probable and stable (while
being evanescent) $(AC)$ "atom", the AC-gas of such "atoms" being
a candidate for dark matter
\cite{5,FKS,Khlopov:2006uv,Belotsky:2006pp}.

However, similar to the sinister Universe AC-lepton relics from
intermediate stages of a multi-step process towards a final $(AC)$
atom formation must survive with high abundance of {\it visible}
relics in the present Universe. In spite of the assumed excess of
particles ($A^{--}$ and $C^{++}$) abundance of frozen out
antiparticles ($\bar A^{++}$ and $\bar C^{--}$) is not negligible,
as well as significant fraction of $A^{--}$ and $C^{++}$ remains
unbound, when $AC$ recombination takes place and most of
AC-leptons form $(AC)$ atoms.  This problem of unavoidable
over-abundance of by-products of "incomplete combustion" is
avoided in AC-model owing to the double negative charge of
$A^{--}$ \cite{FKS,Khlopov:2006uv,Belotsky:2006pp}. As soon as
$^4He$ is formed in Big Bang nucleosynthesis it captures all the
free negatively charged heavy particles. Instead of positively
charged ions, created in the Sinister Universe, the primordial
component of free anion-like AC-leptons $A^{--}$ are mostly
trapped in the first three minutes into a puzzling neutral
O-helium state $(^4He^{++}A^{--})$, with nuclear interaction cross
section, which provides anywhere eventual later $(AC)$ binding. As
soon as O-helium forms, it catalyzes in the first three minutes
effective binding in $(AC)$ atoms and complete annihilation of
antiparticles. Products of annihilation cause undesirable effect
neither in CMB spectrum, nor in light element abundances.
O-helium, this surprising $\alpha$-particle with screened Coulomb
barrier, can influence the chemical evolution of ordinary matter,
but might not result in over-production of anomalous isotopes
\cite{FKS,Khlopov:2006uv,Belotsky:2006pp}.

At small energy transfer $\Delta E \ll m$ cross section for
interaction of AC-atoms with matter is suppressed by the factor
$\sim Z^2 (\Delta E/m)^2$, being for scattering on nuclei with
charge $Z$ and atomic weight $A$ of the order of $\sigma_{ACZ}
\sim Z^2/\pi (\Delta E/m)^2 \sigma_{AC} \sim Z^2 A^2 10^{-43}
\cm^2 /S^2_2.$ Here we take $\Delta E \sim 2 A m_p v^2$ and
$v/c\sim 10^{-3}$ and find that even for heavy nuclei with $Z \sim
100$ and $A \sim 200$ this cross section does not exceed $4 \cdot
10^{-35} \cm^2 /S^2_2.$ It proves WIMP-like behavior of AC-atoms
in the ordinary matter.

However, composite atom-like WIMPs, being the challenge for
underground DM search (in particular with the use of $^3$He
detector), are inevitably accompanied by a nuclear interacting
O-helium component. It is even possible that this component is the
dominant form of the modern dark matter \cite{anutium}. 

\subsection{Possible effect of O-helium in $^3$He}

The terrestrial matter is
opaque for O-helium and stores all its in-falling flux. 
Therefore, the first evident consequence of the proposed scenario is the
inevitable presence of  O-helium in Earth. If its
interaction with matter is dominantly quasi-elastic, the O-helium flux
moves downward to the center of Earth. If O-helium regeneration is not
effective and $\Delta$ remains bound with heavy nucleus $Z$,
anomalous isotope of $Z-2$ element appears. This is the serious
problem for the considered model.

O-helium density can be expressed through the local dark matter
density $\rho_{o}= \xi_{OHe} \cdot \rho_{loc}$ ($ \xi_{OHe} \le 1$), saturating it
at $ \xi_{OHe}=1$. Even at $ \xi_{OHe}=1$ O-helium gives rise to less than 0.1
\cite{anutium,lom,Belotsky:2006pp} of expected background events
in XQC experiment \cite{XQC}, thus avoiding for all $ \xi_{OHe} \le 1$
severe constraints on Strongly Interacting Massive particles SIMPs
obtained in \cite{McGuire:2001qj} from the results of this
experiment. In underground detectors O-helium species are slowed
down to thermal energies far below the threshold for direct dark
matter detection.

Therefore a special strategy in the search for this form of dark
matter is needed. 

\begin{figure}
\begin{center}
\centerline{\epsfxsize=10cm\epsfbox{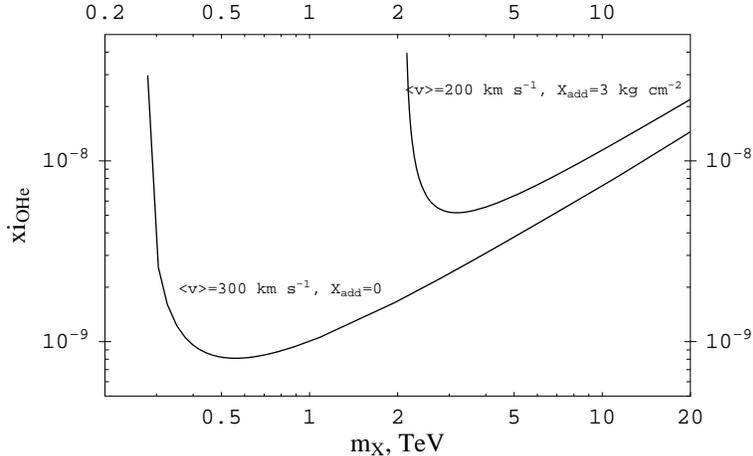}}
\caption{The levels of sensitivity of 1 g of $^3$He detector (corresponding to one event per month of operation) 
to the relative abundance $\xi_{OHe}$ of cosmic OHe in vicinity of Earth versus O-helium mass  $m_X$.
The upper and lower curves correspond to the maximal and minimal estimations of the effect of OHe slowing down. 
The upper curve also takes into account possible effect of slowing down of OHe flux due to collisions with the matter of the roof, walls 
and ceilings of laboratory building.}
\label{fig:OHe}
\end{center}
\end{figure}

An interesting possibility appears with the development of
superfluid $^3$He detectors \cite{Winkelmann:2005un}. Due to the high
sensitivity to energy release above ($E_{th} = 1 \keV$), the operation
of a few gram prototype can put severe constraints on a
wide range of $ \xi_{OHe}$ and O-helium mass $m_X$. 
We can illustrate it by the following simple estimation. Indeed, the initial kinetic energy $E\sim \frac{m_X (300\, {\rm km/s}/c)^2}{2}$
of cosmic O-helium of mass $m_X$,  falling downward to the center of the Earth, decreases due to collisions with matter nuclei 
with the rate per the unit length
\begin{equation}
dE/dX=-\Delta E /X_{mat}.
\label{LossDE}
\end{equation} 
Here $\Delta E=\frac{2m_{A}}{m_X}E$ is the averaged energy loss in a collision with atomic nucleus with the  
mass $m_A$. The mean mass of nuclei in atmosphere can be taken as  $m_A \approx 15$ and the nuclear collision length in it is
$X_{mat}\approx 60$ g/cm$^2$. We can also take into account additional energy losses due to collisions with atomic nuclei 
of matter in roof, ceilings and walls of laboratory building.
To be registered by a He-3 detector, the kinetic energy of slowed down O-helium, which is determined by Eq.(\ref{LossDE}) and given by
\begin{equation}
E_{det}=E\exp\left(-\frac{2m_{A}}{m_X}\frac{X(\theta)}{X_{mat}}\right),
\end{equation} 
should be sufficiently high to produce a signal in the detector above the threshold $E_{th}$.
The length $X(\theta)$ travelled by a cosmic O-helium before reaching the detector depends on its initial direction
and includes the thickness of the atmosphere $X_{atm}(\theta)$ and the additional thickness $X_{add}$ of the laboratory bulding
where the detector operates. The condition that the  energy transfer to $^3He$ exceeds the detection threshold
$E_{Rmax}=4m_{^3He}E_{det}/m_{X}>E_{th}=1$ keV fixes possible directions (solid angles) for incoming O-helium flux
to be detected in the He-3 detector.
The level of sensitivity of 1 g of superfluid $^3He$ to the local relative density of O-helium 
with mass $m_X$ is shown on the figure \ref{fig:OHe}.
This level is roughly estimated as corresponding to one event registered during one month of detector operation. 
The simulation of O-helium events in the detector and the analysis of their detection efficiency are of special interest
and will be considered elsewhere.

These estimations demonstrate that the analysis of the data from the existing detector prototype can provide a sensitive 
test of composite dark matter models in a wide range of their parameters. O-helium with smaller masses can escape this 
test due to effective slowing down below the detection threshold. However, at these masses experimental search for charged
consituents of O- helium ($A$-leptons of AC-model or stable quarks of 4th generation of model \cite{anutium}) 
is possible in the nearest future at accelerators \cite{Belotsky:2006pp}.

\section{Superfluid He-3 test for Dark Matter}

\subsection{Superfluid $^3$He at ultra-low temperatures}
Many collaborations have developed promising detectors to search for non-baryonic dark matter.
These detectors have reached sufficient sensitivity to probe the existence of various dark matter candidates,
including even sparse sub-dominant components. In particular, such sensitivity allows to test some regions
of the SUSY parameter space \cite{machet-plb} or properties of 4th family of quarks and leptons \cite{Fargion}.

Direct detection experiments present common problems such as neutron interaction background and
 radioactivity contamination from both the sensitive medium and the surrounding materials. Substantial
experimental research has been devoted to the development of different types of detectors, in order to optimize
the sensitivity while keeping to a minimum the undesirable effects. Based on early experimental works, a superfluid
$^3$He detector for direct detection of non-baryonic dark matter has been proposed \cite{First, Second}.
The first experimental tests of a $^3$He detector by neutrons and $\gamma$-rays have been done in Lancaster and
Grenoble  \cite{BOLO,Gren2,Gren}.
\begin{figure}
\begin{center}
\centerline{\epsfxsize=12cm\epsfbox{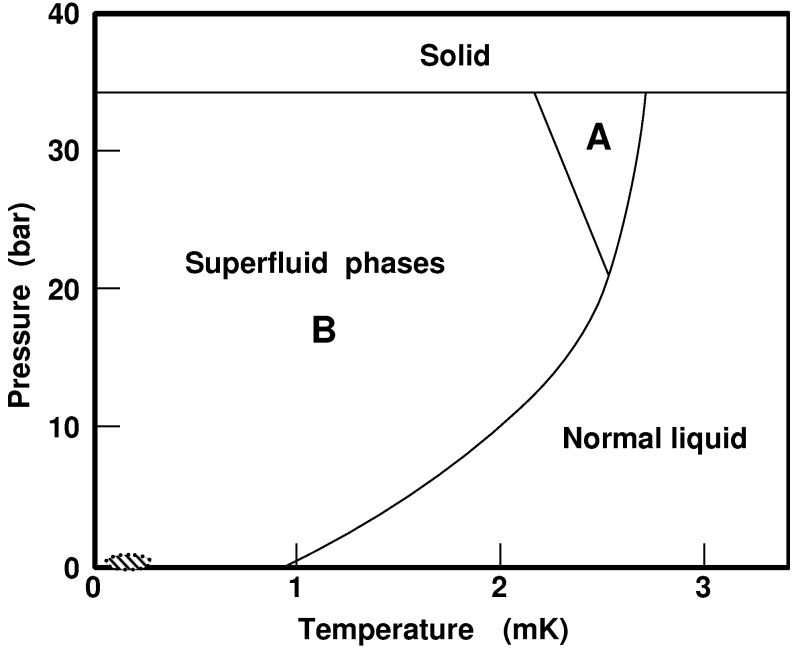}}
\caption{$^3$He phase diagram in the milliKelvin range and for 0 magnetic field.
Above 34\,bar $^3$He is solid at these temperatures. The ``normal" Fermi liquid $^3$He has a superfluid phase
transition at a temperature $T_c$ varying from 0.93 to 2.49\,mK between vapor pressure and melting pressure.
The complex order parameter associated to the phase transition allows the existence of various superfluid phases
with different broken symmetries, two of which are represented here (see text for details). The working point of the
detector is within the shaded area near the origin.}
\label{fig:phase-diagram}
\end{center}
\end{figure}

$^3$He is a quantum fluid obeying Fermi statistics, and it remains liquid down to the absolute zero of temperature.
At about 1$\sim$2.5 mK (depending on the pressure), liquid $^3$He displays a second order phase transition
to its superfluid A- and B-phases, as shown in figure \ref{fig:phase-diagram}.
The superfluid A phase has an anisotropic gap structure and an order parameter mixing magnetic
and flow properties, while the B phase is characterized by an isotropic gap $\Delta=1.76\ k_BT_c$ well described at 0 pressure
by the weak coupling BCS theory \cite{leggett}.
Experimental temperatures as low as 100 $\mu$K are achieved by adiabatic nuclear demagnetization of a copper stage,
which then cools down the liquid $^3$He \cite{pobell}. At these temperatures far below the transition temperature $T_c$,
the superfluid is in its isotropic B-phase and the density of thermal excitations (quasiparticles) $n$
decreases exponentially with temperature
\begin{equation}
 n=\int g(E)dE = \frac{N_A}{V}\sqrt{ 2\pi\frac{\Delta}{k_BT}}\ \exp(-\Delta/k_BT),
\label{eq:density}
\end{equation}
where $g(E)$ is the density of states, $N_A$ is Avogadro's number and $V$ the molar volume of the fluid.
This density is so low that the liquid can be represented as a renormalized quantum vacuum carrying a dilute quasiparticle gas.
In the range of 100 to 200$\ \mu$K, the heat capacity of the superfluid is dominated by the ballistic quasiparticle gas and reduces to
\begin{equation}
C=C_0 \left(\frac{T_c}{T}\right)^{3/2}\exp(-\Delta/k_BT),
\label{eq:heat-capa}
\end{equation}
with $C_0\approx2.1$ mJ K$^{-1}$cm$^{-3}$.

A direct and rather rapid method of thermometry of the superfluid is achieved by measuring the density of thermal excitations
(quasiparticles) using Vibrating Wire Resonators (VWRs) \cite{thermo}. A VWR is a fine superconducting wire
bent into semi-circular shape and oscillating perpendicularly to its plane. The excitation and the read-out of the
VWR are obtained by electrical (a.c. current) means.

The dynamics of the VWR can be conveniently described by a damped harmonic oscillator model.
The damping of the VWR is dominated by the friction with the quasiparticle gas of the surrounding superfluid~\cite{lancaster}.
The damping of the oscillator and thus the resonance line-width at half-height $W$
are proportional to $\int v_g\ g(E)dE $, where $v_g$ is the quasiparticle group velocity, and may be expressed
as a function of temperature
\begin{equation}
W(T)=W_0\  \exp(   -\Delta/k_BT).
\label{eq:heat-capa}
\end{equation}
$W$ exhibits therefore an extremely steep temperature dependence around 100\,$\mu$K.
The continuous measurement of $W$ allows to access temperature with response
times $\tau_{wire}=1/\pi W < 1$ s for typical values of $W$ of the order of 1\,Hz. The VWR being immersed
in the target medium provides a sensitive and direct measurement of the temperature. 

\subsection{$^3$He as a target material for Dark Matter search}

For the bolometric particle detection we use copper cells of typical dimensions about 5 mm, filled with superfluid
$^3$He which is in weak thermal contact with the outer bath through a small orifice \cite{BOLO, Gren2,bolo-jltp}.
The interaction of a particle with the $^3$He in the cell releases energy which results in an increase of temperature, and thus $n$.
The time constants of internal equilibrium of the quasi-particle gas are small ($<$ 1 ms), while the time constant for thermal
relaxation of quasiparticles through the orifice after a heating event is tuned to be $\tau_{cell} \approx 5$ s.
The heat leak through the container walls can be neglected because of the huge thermal resistance (Kapitza resistance)
of solid-liquid interfaces at very low temperatures. Each bolometric cell contains a least one VWR-thermometer which allows
to follow the rapid variations of the temperature.

Bolometric calibration of the detector cells is achieved by an extra VWR present in the cell that can produce
a short mechanical pulse at its resonant frequency and thus deposit a well-controlled amount of energy (heat)
to the liquid through mechanical friction \cite{Gren2, bolo-jltp}.

While the use of $^3$He imposes challenging technological - namely cryogenical - constraints,
this material has nevertheless extremely appealing features for Dark Matter detection~:\\
\indent{$\bullet$} The $^3$He nucleus having a non-zero magnetic moment, a $^3$He detector will
be mainly sensitive to the axial interaction \cite{machet-plb}, making this device complementary to
existing ones, mainly sensitive to the scalar interaction. The axial interaction is largely dominant in most of
the SUSY region associated with a substantial elastic cross-section.\\
\indent{$\bullet$} The purity of bulk liquid $^3$He at 100 $\mu$K is virtually absolute.
Nothing can dissolve in $^3$He at these temperatures; no magnetic impurities.\\
\indent{$\bullet$} $^3$He presents the rather unique feature of a high neutron capture cross-section. The nuclear capture reaction
\begin{equation}
{\rm n\ +\ ^3He\ \rightarrow\ p\ +\ ^3H}
\label{eq:capture}
\end{equation}
leads to a large energy release of 764 keV. Neutron contamination has thus a clear signature~\cite{BOLO,Gren},
well discriminated from a WIMP signal. 

\indent{$\bullet$} The discrimination of elastic neutron scattering can be done by
 well known method of multicell detectors, as it was discussed for superfluid $^3$He case in \cite{machet-nim}.
 The multiscattering of neutron in a few cells can be considered as a veto. 
Owing the liquid nature of $^3$He the dead volume between the cells can be so small as a microns copper foils of the 
walls between the cells. 
Consequently, the probability for neutron to escape after a single scattering can be minimized.
\\
\indent{$\bullet$} A high transparency to $\gamma$-rays due to low Compton cross-section and the absence of photoelectric effect.
No intrinsic X-rays.\\
\indent{$\bullet$} A high signal to noise ratio, due to the narrow energy integration range expected for a WIMP signal.
Since the target nucleus (${\rm m=2.81 \,GeV}\!/c^2$) is much lighter than the incoming  neutralino,
the maximum recoil energy does only depend weakly on the WIMP mass. As a matter of fact, the recoil energy range
needs to be studied only below ${\rm 6 \,keV}$~\cite{machet-plb,machet-nim}.\\
\indent{$\bullet$} The target material being a quantum liquid, it has no coherent recoil unlike crystals and can be most easily recycled.

\section{The current state: neutron, muon and low energy electron test of the superfluid $^3$He bolometer}

\subsection{Neutrons}

\begin{figure}
\begin{center}
\centerline{\epsfxsize=12cm\epsfbox{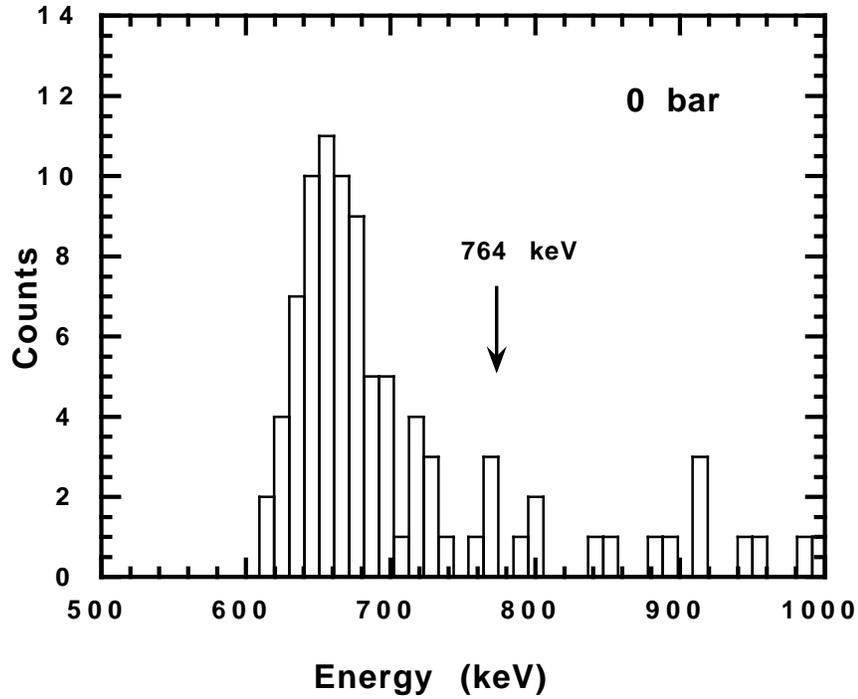}}
\caption{Detected energy spectrum associated to neutrons.
An independent bolometric calibration of the cell allows to compare quantitatively the measured peak to the expected 764 keV,
yielding a deficit of about 120 keV (see text).}
\label{fig:neutrons2}
\end{center}
\end{figure}

Neutrons were the first particles studied in superfluid $^3$He \cite{BOLO} for their large energy release (\ref{eq:capture}).
Of particular interest was the study of the rapid and inhomogeneous phase transition of a small region around the neutron impact,
for the possibility of topological defect creation in the superfluid in analogy with the Kibble mechanism in cosmology \cite{Gren}.
The neutrons, emitted by a moderated AmBe source, produce large signals in the bolometer.
A deficit of about 120 keV with respect to the expected 764 keV is observed (figure \ref{fig:neutrons2});
part of this deficit is accounted for by ultra-violet scintillation of the $^3$He, the rest is interpreted in terms of energy
trapped in the form of metastable topological defects of the superfluid (e. g. quantized vortices).

\subsection{Muons}

\begin{figure}
\begin{center}
\centerline{\epsfxsize=12cm\epsfbox{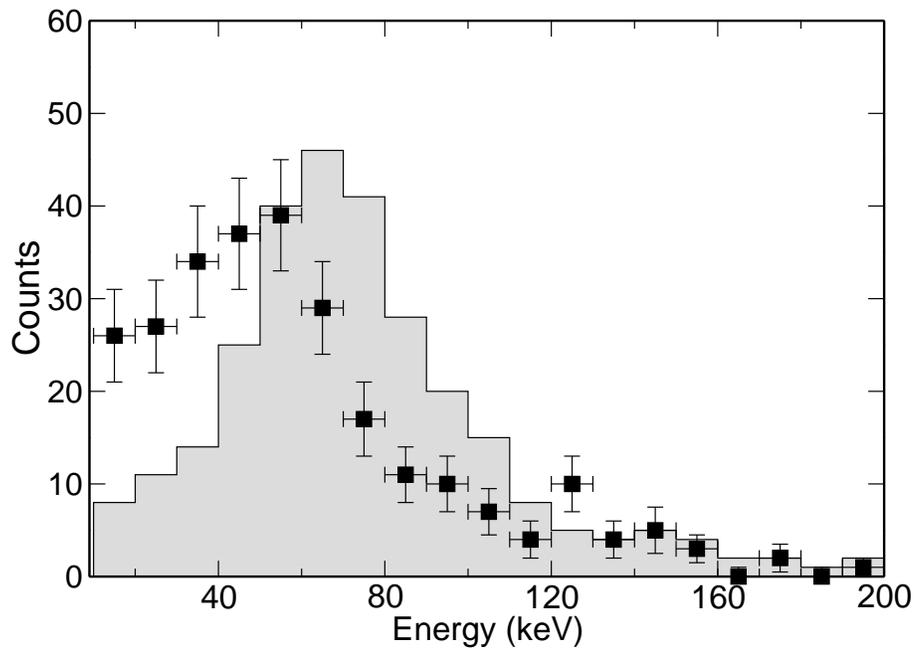}} 
\caption{Cosmic muon detection spectrum, calculated (histogram)
and experimental (points). The detection spectra qualitatively agree; the numerical simulation calculating the energy
release of the muons without taking into account losses by ultra-violet scintillation, the difference of the maxima of about
25\% allows to estimate the scintillation rate produced by ionizing events in superfluid $^3$He.}
\label{fig:muons1}
\end{center}
\end{figure}

Cosmic muons are expected to deposit about 16 keV/mm in liquid $^3$He at 0 bar.
Muons represent thus bolometric events about an order of magnitude below neutrons.
A muon test of the detector and its comparison to a numerical simulation by Geant4 in the frame of the MACHe3 collaboration
yielded good agreement (figure \ref{fig:muons1}), the 20-25\,\% difference between the experimental and
the calculated detection spectra being due to to ultra-violet scintillation \cite{eddy}. Both peaks are smoothed through
the geometrical averaging over all possible incidence angles.

Final evidence for the muonic nature of the observed energy peak at about 50$\sim$60 keV at ground level was brought
by the recent experiment with a 3-cell prototype. The simultaneous detection in 3 adjacent cells allowed to discriminate
with large efficiency the muons, who are, depending on their trajectory, generally detected coincidently in two or more cells.
This setup therefore allowed to demonstrate the large muon rejection efficiency of a future underground multicell detector.

\subsection{Electrons}

\begin{figure}
\begin{center}
\centerline{\epsfxsize=12cm\epsfbox{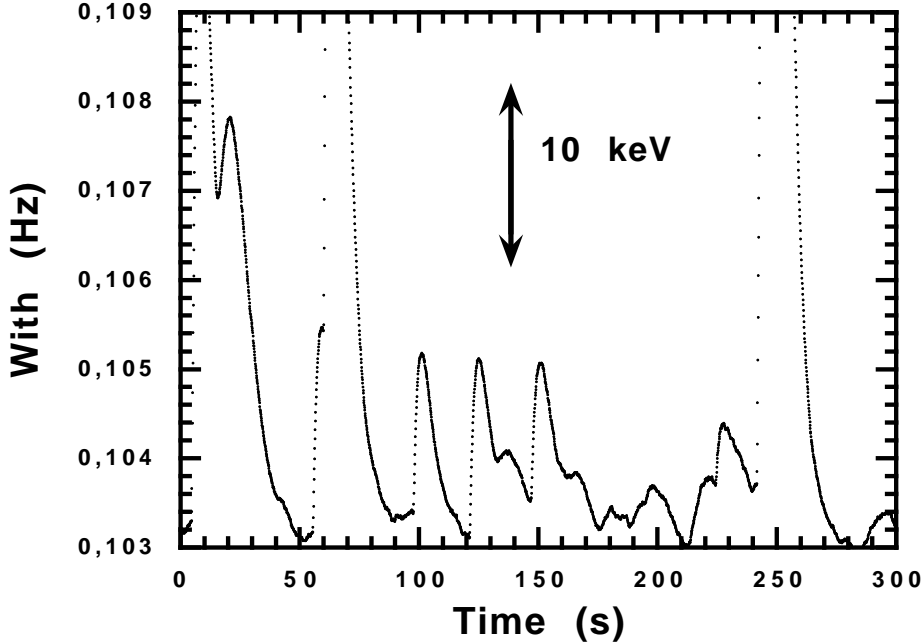}} 
\caption{Detection of single electrons at keV level
from the $^{57}$Co source. Events at about 7 and 14 keV are easily detected.
Even some unidentified bolometric events as small as 1 keV are clearly visible.}
\label{fig:electrons1}
\end{center}
\end{figure}

\begin{figure}
\begin{center}
\centerline{\epsfxsize=12cm\epsfbox{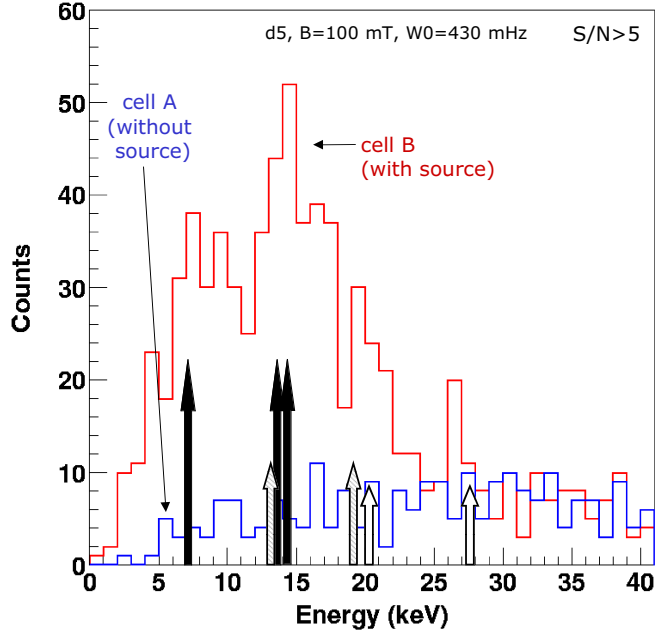}} 
\caption{Detection spectrum in cell B (with source) at low energies.
The observed spectrum coincides very well with the expected electron lines from the source (arrows). For comparison,
the spectrum in cell A (without source) does not display any comparable structure (arrow shows the data from second cell).
Note that the energy scale chosen takes into account the ultra-violet losses of about 25\,\% of ionising event.}
\label{fig:electrons2}
\end{center}
\end{figure}

Since the energy range of a neutralino scattering is expected to be in the keV range, the proof that
a 1\,keV detection resolution and threshold could be attained had to be brought using a known particle source.
A low activity $^{57}$Co source was therefore implemented directly in one cell (cell B). Such a source emits
$\gamma$-rays mainly at about 120 keV, which have a weak Compton scattering cross-section with the $^3$He,
but also low energy electrons (from internal conversion and Auger effect) which thermalize completely in the liquid of cell B.
Such low energy electron events are expected mainly at about 7 and 14 keV, and only in cell B \cite{moriond,manu}.
Measurements on the 3-cell prototype indeed allowed to identify such bolometric events, again an order of magnitude
below typical muons (figure \ref{fig:electrons1}). The low energy detection spectrum in cell B and its comparison to cell
A (without source) allows to clearly identify these events as produced by the $^{57}$Co source (figure \ref{fig:electrons2}).

\section{Discussion and outlook}
The nature of cosmological dark matter involves physics, going beyond 
the Standard model of electroweak and strong interactions.
Various candidates for dark matter follow from different arguments 
of particle theory and can co-exist in multicomponent dark matter
scenarios. Discrimination of dark matter components inevitably involves 
combination of experimental methods and extension of such methods
of dark matter search is important.

Recently possible set of dark matter candidates was enriched by
idea of composite dark matter, which consists of atom-like systems 
of heavy electrically charged particles. These atom-like systems can have very small size
and behave as weakly interactiing particles. However, it turned out that
realistic scenario of composite dark matter inevitably predicts existence  
of O-helium component (an "atom", in which heavy particle with charge -2
is bound with $^4$He nucleus). This nuclear interacting component can be even 
dominant form of dark matter, but still remain elusive for the existing   
means of direct search for Weakly and Strongly Interaqcting Dark Matter Particles.
New approach to cover the gap in experimental maintainance is needed
to probe the existence of this exotic dark matter component.

$^3$He is a promising target material, though very young in the
particle detection community. The bolometric experiment requires
extremely low working temperatures and superfluid $^3$He is all
but a standard detector material for the search of non-baryonic
Dark Matter. Since the original proposal of the use of $^3$He for
particle detection, the detection threshold and sensitivity have
been improved by 2 orders of magnitude, reaching nowadays 1 keV,
which is already the expected energy range for a neutralino
impact. The use of a $^{57}$Co source producing a well known
$\gamma$-ray and low energy electron spectrum directly in one
bolometric cell allowed to illustrate both our understanding of
the detector at keV level and the high transparency of the target
material to $\gamma$-rays. In addition, the simultaneous detection
in 3 adjacent cells demonstrated the future rejection efficiency
versus ionizing events of a large multi-cell detector. Even the
existing ground-based prototype can shed light on the possible
existence of nuclear interacting DM component (O-helium) in a rather wide
range of its parameters. Analysis of data from this prototype
can provide complete test of composite dark matter models,
being combined 
with future accelerator and cosmic ray searches for charged constituents of
such forms of dark matter.
The following mayor steps are planned to improve the ULTIMA experiment in the coming years:\\

In the absence of artificial laboratory sources, the detection signal is currently largely dominated by cosmic muons.
The next steps will therefore necessarily lead the experiment to an underground laboratory with a muon flux reduced
by 5 or 6 orders of magnitude.\\

Parallel methods of discrimination of ionizing events are in study.
As in other materials, the fraction of energy emitted by the $^3$He as ultra-violet scintillation can provide 
a fine criterium of discrimination.\\
Alternative methods of thermometry to the classical VWR are currently thought of. Microfabricated silicon VWRs
are already being produced and tested in Grenoble \cite{seb}
and will allow to mass-fabricate the bolometric  cells. In parallel, a
method of thermometry based on Nuclear Magnetic Resonance (NMR) of
the superfluid is currently being tested and may provide in the
future a thermometric probe at 100\,$\mu$K
with time constants below 1\,ms.\\

 The increase of the target mass to about 100 g should be possible in the coming years without demanding
any extreme technological and cryogenical advance. As we have shown in the present paper 
such a target mass would already allow
to test the current positive results of the DAMA experiment, sensitive to the spin-dependent interaction \cite{DAMA-review}.
Depending on the success of the experiment, a more challenging increase of the target mass to 1$\sim$10 kg may be thought of.
Such a large detector using 10$^3$ independent bolometric cells would allow to look deep into the parameter space
of possible Dark Matter models.\\

\section{Acknowledgments}
This work was partially funded by the R\'egion Rh\^one-Alpes, by the
European Community under the Competitive and Sustainable Growth
programme (Contract G6RD-CT-1999-00119), by the Bureau National de
M\'etrologie (Contract 00 3004). The development of the new
$^3$He detector is supported by the Agence Nationale de la Recherche
(France). M.Kh. thanks CRTBT-CNRS and LPSC, Grenoble for
hospitality. The work of M.Kh. and K.B. was partially supported by 
the State grant for Russian Scientific Khalatnikov-Starobinsky school.
The authors wish to thank D. Fargion, V. Goudon and E. Collin for
their precious help.

\end{document}